\documentstyle[twoside,fleqn,espcrc2,epsf]{article}

\newcommand{\bi}{\begin{itemize}}
  \newcommand{\ei}{\end{itemize}}

\newcommand{\beq}{\begin{equation}}
  \newcommand{\eeq}{\end{equation}}

\newcommand{\PR}{Phys.\ Rev.~D }
\newcommand{\PRL}{Phys. Rev. Lett.}

\newcommand{\PL}{Phys.\ Lett.}

\newcommand{\bea}{\begin{eqnarray}}
\newcommand{\ena}{\end{eqnarray}}

\title{Improving Stochastic Estimator Techniques for 
Disconnected Diagrams\thanks{Talk presented by J. Viehoff.}}
\author{{\sf SESAM}-Collaboration: J.~Viehoff$^{\rm b}$,
                                   N.~Eicker$^{\rm a}$, 
                                   S.~G\"usken$^{\rm b}$, 
                                   H.~Hoeber$^{\rm b}$, 
                                   P.~Lacock$^{\rm a}$, 
                                   Th.~Lippert$^{\rm a}$, 
                                   G.~Ritzenhoefer$^{\rm a}$,
                                   K.~Schilling$^{\rm a,b}$,
                                   A.~Spitz$^{\rm a}$ and
                                   P.~Ueberholz$^{\rm b}$ \\[8pt]
{\small  {\rm $^a$}HLRZ c/o Forschungszentrum J\"ulich, D-52425 J\"ulich,
  and DESY, D-22603 Hamburg, Germany,\\
  {\rm $^b$}Physics Department, University of Wuppertal, D-42097
  Wuppertal, Germany.}}       
\begin{document}
\begin{abstract}
Disconnected diagrams are expected to be sensitive to the inclusion of
dynamical fermions. We present a feasibility study for the observation
of such effects on the nucleonic matrix elements of the axial vector
current, using SESAM full QCD vacuum configurations with Wilson fermions on
$16^3\times 32$ lattices, at $\beta =5.6$. Starting from the standard methods developed
by the Kentucky and Tsukuba groups, we investigate the improvement from
various  refinements thereof.
\end{abstract}
\maketitle
\section{Introduction\label{INTRO}}
The calculation of hadronic matrix elements containing flavor singlet 
quark bilinear operators ${\cal O}=\bar{q}\Gamma q$ is expected to reveal 
sea quark effects in full QCD simulations. 

The standard method for such calculations uses the ratio of the three- and the 
two-point functions:
\bea
R(t)&=&\frac{\langle N(t)\sum_n{\cal O}(n)\bar{N}(0)\rangle}{\langle N(t)\bar{N}(
0)\rangle}\nonumber\\
\nonumber\\
&&\stackrel{t\gg 1}{\longrightarrow} const +Z_0^{-1}\langle N | {\cal O}| N\rangle \,t\; ,
\label{ratio}
\ena
where $Z_0$ is the lattice renormalisation factor for ${\cal O}$.
In the limit of large $t$ $R(t)$ shows a constant slope, which determines the 
matrix element under investigation. In this work we focus on scalar and axial vector insertions 
($\Gamma =1$, $\Gamma = \gamma_k\gamma_5$) needed for the determination of the
$\pi N\sigma$ term, which measures chiral symmetry breaking and for the axial 
coupling constant, $g_A^1$. In both cases connected and disconnected
diagrams have to be dealt with. For the former suitable methods are available. However, the latter
present a severe bottleneck in form of the computation of
$\Delta =(\Gamma M^{-1})_{x\to x}$ which is of complexity ${\cal O}$(volume)\cite{Mandula}.

\section{Improving Scalar Insertions}
To tackle the bottleneck several estimator techniques have been devised by different groups
\cite{Kentucky1,Kentucky2,Fukugita}. 
All these methods approximate $\Delta$ by scalar products $(\eta , \Gamma M^{-1}\eta)$  with
appropriate source vectors $\eta$. In the so called wall source 
method a homogeneous $4$-volume source is used\cite{Fukugita}, leading to unbiased 
observables only in the limit of an infinite ensemble of gauge configurations. 
In view of the cost of sampling this poses a non trivial problem for full QCD simulations.

This problem is alleviated by the stochastic estimator techniques (SET)\cite{Kentucky1,Kentucky2}.

It has been shown previously\cite{discon} that a complex $Z_2$ noise source is superior to Gaussian noise,
in the case of the scalar insertion. In this contribution we present new methods to improve
on the signal to noise ratio for scalar and axial vector insertions.  

One possible strategy to reduce the noise is to restrict the summation over $n$ in 
eq. \ref{ratio} to the plateau region of the proton correlator, i.e. to the ground state
regime of the proton operator. 
We refer to this approach as PSQL (plateau sampling of quark loops). 

\begin{figure}[htb]
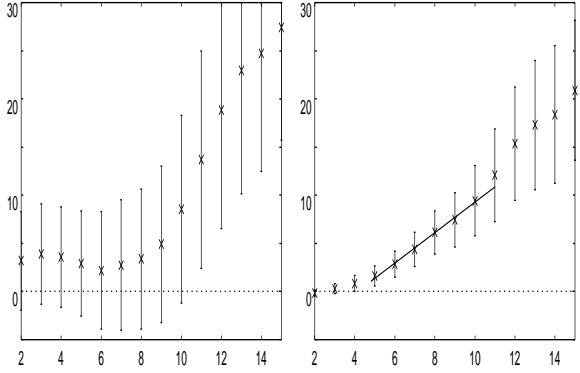

\begin{center}
\noindent\parbox{7.6cm}{
\parbox{3.8cm}{\epsfxsize=3.7cm\epsfysize=5cm\epsfbox{plot4.eps}}\nolinebreak\hfill
\parbox{3.8cm}{\epsfxsize=3.7cm\epsfysize=5cm\epsfbox{plot5.eps}}
}
\end{center}
\vskip -0.3cm
\caption[a]{\label{scalar}
\it $R(t)$ for the scalar insertion measured without PSQL (left) and with PSQL 
on $200$ SESAM configurations, $100$ $Z2$ noise estimates, $\kappa =0.157$.} 
\end{figure}

The efficiency of PSQL (with $1\le n_t\le t-1$) is illustrated in figure \ref{scalar}, where we 
compare with standard SET. Note that the signal to noise ratio is improved by a factor $2-3$.  

\section{Axial insertions}
The matrix element of the axial vector current between proton states (for
zero momentum) can be written as:
\bea
\langle P| \bar{q_i} \gamma_{\mu}\gamma_5q_i | P\rangle = 
g^i_A \;\bar{P} \gamma_{\mu}\gamma_5 P \; , \; i=u,d,s.
\ena
The quantity of interest is the flavor singlet coupling
\bea
g_A^1=\sum_{i=u,d,s}{g_A^i}.
\ena

\begin{figure}[htb]
\begin{center}
\noindent\parbox{7.6cm}{
\parbox{7cm}{\epsfxsize=7cm\epsfysize=5cm\epsfbox{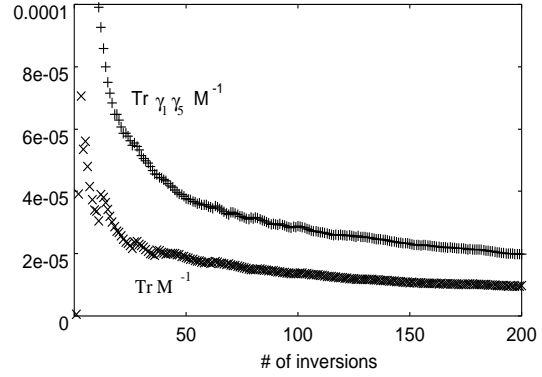}}}
\end{center}
\vskip -0.7cm
\caption[a]{\label{standerr}
\it The standard error $\sigma /\sqrt{N}$ on one configuration for $Tr(M^{-1})$ and 
$Tr(\gamma_1\gamma_5M^{-1})$ with Z$2$ noise source vectors.}
\end{figure}

To determine $g_A^1$ on the lattice one needs to compute 
$Tr(\gamma_k\gamma_5M^{-1})$ which implies off diagonal elements of $M^{-1}$ in
spin space\cite{axial}. 
Unfortunately the SET becomes inefficient for such matrix elements, as illustrated in figure
\ref{standerr}. 
To improve the situation we resort to an explicit, component-wise treatment in spin space.
This amounts to the application of one independent stochastic inversion per spin component
(spin explicit method, SEM). We emphasize that SEM can be combined with PSQL. 

\begin{figure}[htb]
\begin{center}
\noindent\parbox{7.6cm}{
\parbox{7cm}{\epsfxsize=7cm\epsfysize=7cm\epsfbox{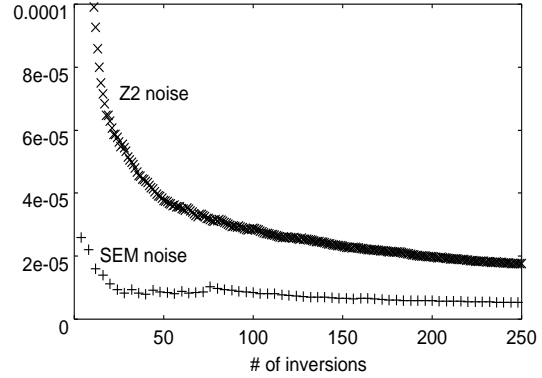}}}
\end{center}
\vskip -2.7cm
\caption[a]{\label{standerr2}
\it The standard error $\sigma /\sqrt{N}$ on one configuration for $Tr(\gamma_1\gamma_5M^{-1})$
for Z2 and SEM noise source vectors.}
\end{figure}

\begin{figure}[htb]
\begin{center}
\noindent\parbox{7.6cm}{
\parbox{7cm}{\epsfxsize=7cm\epsfysize=5cm\epsfbox{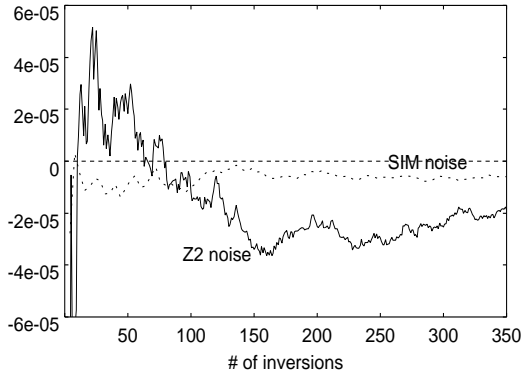}}}
\end{center}
\vskip -0.7cm
\caption[a]{\label{meanvalue}
\it Mean values for $Tr(\gamma_1\gamma_5M^{-1})$ on one configuration for Z2 and 
SEM noise source vectors.}
\end{figure}

The effect is illustrated in figure \ref{standerr2} and \ref{meanvalue} for the standard error and
the mean values on a single configuration. We observe earlier asymptotics in terms of
the number of inversions (the factor $4$ for SEM has been taken into account).

\begin{figure}[htb]
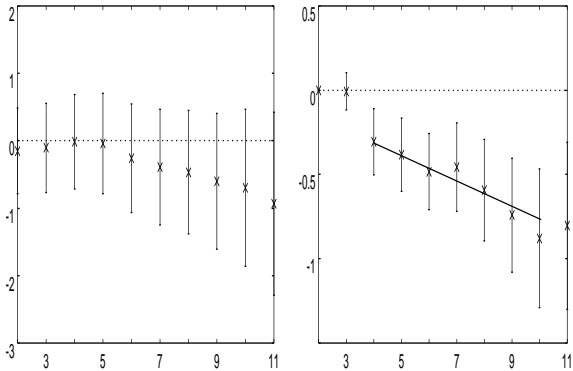

\begin{center}
\noindent\parbox{7.6cm}{
\parbox{3.8cm}{\epsfxsize=3.7cm\epsfysize=5cm\epsfbox{plot3.eps}}\nolinebreak\hfill
\parbox{3.8cm}{\epsfxsize=3.7cm\epsfysize=5cm\epsfbox{plot2.eps}}
}
\end{center}
\vskip -0.3cm
\caption[a]{\label{g_A}
\it $R(t)^{disc}_{g_A}$ measured without PSQL (left) and with PSQL on $200$ 
SESAM configurations, $100$ SEM noise source vectors, $\kappa =0.157$.} 
\end{figure}

In order to test the sensitivity of the method for a determination of $g_A^1$ 
we have analysed $200$ SESAM configurations, at $\kappa = 0.157$.
From the ratio 
\begin{eqnarray}
R(t)^{disc}_{g_A} &=& \frac{ \langle P_{11}(0 \rightarrow t) Tr(\gamma_3\gamma_5M^{-1}) \rangle}
                {\langle P_{11}(0 \rightarrow t)\rangle } \nonumber \\
 & & - \langle Tr(\gamma_3\gamma_5M^{-1})
             \rangle
\nonumber \\ \nonumber \\
             \stackrel{t \; large}{\longrightarrow}
             &\mbox{const}& + t\,\langle P_1|\bar{q}\gamma_3\gamma_5q|P_1\rangle^{latt}_{disc} \; 
\end{eqnarray}
we obtain signals as given in figure \ref{g_A}. Using both, SEM and PSQL, one finds the signal 
to stick clearly out of the noise!  

\section{Conclusion and outlook\label{conclusion}}
We have presented a variant of the SET which is suited to measure the flavor singlet 
axial vector coupling of the nucleon on a limited sample of QCD configurations. We can 
cope with the smallness of the axial vector
signal by spin explicit inversions. Although this amounts to an increase in computational 
cost by a factor $4$ for each estimate, the overhead turns out to be largely compensated
by earlier asymptotics. The second ingredient of our proposal is the use of plateau
sampling, which appears to reduce the fluctuations by a factor $2-3$, both for the scalar
and for the axial insertions. 

This encourages us to perform a full physics analysis to the complete set of SESAM configurations.


\begin{thebibliography}{99}
\bibitem{Mandula}{J.E. Mandula, M.C. Ogilvie, \PL {\bf B312}(1993)327.}
\bibitem{Kentucky1}{S.J. Dong and K.F. Liu, Nucl.Phys. {\bf B} \break (Proc.Suppl.)
{\bf 26} (1992) 353; \PL {\bf B328} (1994) 130.}
\bibitem{Kentucky2}{S.J. Dong and K.F. Liu, Nucl.Phys. {\bf B} (Proc.Suppl.)
{\bf 26} (1993) 487; K.F. Liu, S.J. Dong, T. Draper
and W. Wilcox, \PRL {\bf 74} (1995) 2172; S.J. Dong, J.F. Laga\"e, and
K.F. Liu, \PRL {\bf 75} (1995) 2096.}
\bibitem{Fukugita}{ Y. Kuramashi, M. Fukugita, H. Mino, M. Okawa, and
A. Ukawa,  \PRL {\bf 75} (1995) 2092; \PR {\bf 51} (1995) 5319.}
\bibitem{axial}{S.J. Dong, J.-F. Lagae, K.F. Liu, Phys.Rev.Lett. {\bf 75} (1995) 2096-2099.}
\bibitem{discon}{SESAM Collaboration: N.~Eicker, S.~G\"usken, H.~Hoeber, P.~Lacock,
Th.~Lippert, K.~Schilling, A.~Spitz, P.~Ueberholz and J.~Viehoff, Phys.Lett. 
{\bf B389} (1996) 720-726.}
\end{thebibliography}
\end{document}